# Modified Relational Quantum Mechanics

**B.K. Jennings**

A modified version of relational quantum mechanics is developed based on the three following ideas. An observer can develop an internally consistent description of the universe but it will, of necessity, differ in particulars from the description developed by any other observer. The state vector is epistomological and relative to a given quantum system as in the original relational quantum mechanics. If two quantum systems are entangled, they will observe themselves to be in just one of the many states in the Schmidt biorthonormal decomposition and not in a linear combination of many.

1. Introduction

In his original paper on relational quantum mechanics[1] Rovelli suggested that its relation to Breuer's work[2] on the impossibility of an observer to distinguish all present states of a system in which he or she is contained should be explored. This work takes that suggestion seriously and comes to the conclusion that any entangled system necessarily leads to wave function collapse.

Rovelli stated[1]: "Quantum mechanics is a theory about the physical description of physical systems relative to other systems, and this is a complete description of the world." This is modified to: "Quantum mechanics is a theory about the physical description of physical systems relative to other systems *and to itself*, and this is a complete description of the world".

The discussion begins with a few general observations. Facts are relative to a given observer (quantum system)[1] and are the result of a measurement. A measurement occurs with respect to the quantum system B when, through an interaction, it becomes entangled with quantum system A such that a set of initial properties of A are unambiguously mapped to final states of B. It is always possible to acquire new information about a system by making a measurement. As a corollary a measurement causes the loss as well as the gain in information[1]. The use of the term "facts" has no ontological implication. In this regard the term "observation" is preferred for the result of a measurement. The pragmatic philosopher Clarence I. Lewis (1883 - 1964) advocated that observations are, in general, relative to the observer, see chapter 6 of reference 3.

Quantum mechanics is based on probability amplitudes and not on straight probabilities. The amplitude is frequently referred to as the state of the system or its wave function. The amplitude of a given system evolves by a unitary transformation relative to an observer not interacting with the given system. Probability amplitudes have phases as well magnitudes. Decoherence is the loss of phase information through the effective contraction of the reduced density matrix to a diagonal form. Decoherence is different from wave function collapse. The first leaves the reduced density matrix diagonal but with possibly many terms. The second reduces the true (with respect to a given observer) density matrix to a single diagonal term.

2. **The Schmidt Biorthonormal Decomposition**.

Any entangled state of two systems can be expanded in a Schmidt biorthonormal decomposition[4], the wave function being written as $\Psi(A,B) = \sum_i c(i) |A(i)>|B(i)>$ where the $|A(i)>$ and $|B(i)>$ are orthonormal bases in the two spaces A and B respectively. The basis states are determined solely by the entangled state. The $c(i)$ are complex coefficients. The decomposition is unique except when two or more coefficients have equal magnitudes. Then there can be any linear combination in the degenerate space. There is also a trivial ambiguity about how the phase of a single term is distributed between the two states and the expansion coefficient. This has no observable or meaningful consequences.

The trace over one of an entangled pair of quantum systems always leaves the reduced density matrix of the other one diagonal. If a third system does not interact with system A, the reduced density matrix for B relative to the third system can be written as $\rho = \sum_i |B(i)> |c(i)|^2 <B(i)|$. This is decoherence. Since there is no interaction the wave function does not change. Wave function collapse is the reduction of the entangled state to one term in the Schmidt biorthonormal decomposition: $\Psi(A,B) = \sum_i c(i) |A(i)>|B(i)> \implies |A(j)>|B(j)>$. The collapse is only relative to A or B.

Thomas Breuer[2] showed that neither A nor B can observe (measure) the phases of the $c(i)$. To apply to quantum mechanics this relies on quantum mechanics being based on probability amplitudes. Thus the arguments do not apply unmodified to the de Broglie-Bohm interpretation of quantum mechanics.

If we break the universe into two systems, an observer and the rest of the universe, the phases that can not be measured are observer dependent, depending on how the observer system is chosen. Thus the description of the universe and by implication the quantum probability amplitudes are relative to a given observer.

Relative to observers A or B, their entanglement can not be due to unitary time evolution since if it was they would know the phases. The evolution would be unitary with respect to an independent third system. A and B must both observe that they are in a pure state corresponding to one term in the biorthonormal expansion since only the pure state does not depend on the unknowable phases. For an independent third system the wave function is still fully entangled.

Wave function collapse is due to the entangled systems "knowing" which state in the Schmidt decomposition they are in. Note that the collapse is only with respect to the given observers A and B and leaves their density matrix with one diagonal term. If we interpret the wave function epistomologically it would be expected that self observation would result in a system knowing what state it is in leading to wave function collapse relative to that system.

Rovelli's claim[1] that the breakdown of unitary time evolution is because B cannot know its own interactions provides a simple explanation for the lack of unitary evolution Breuer's result implies. To fully explain wave function collapse requires **in addition** that the system **observes itself** to be in a single state in the Schmidt decomposition after its unknown interaction. This will be referred to as self observation. Alternately one could say B's observation of itself causes the wave function to collapse, the time evolution to be non unitary and the interaction leading to the entangled state to be ill defined, all relative to B.

The key point is that self observation naturally leads to wave function collapse and the Schmidt decomposition gives a unique set of states into which it can collapse. Breuer showed[2] why the Schmidt decomposition is special. The self observation discussion should be regarded as heuristic rather than rigorous. However in relational quantum mechanics, self observation can be assumed without contradiction as a separate axiom.

3. **The Measurement Process**

In a typical measurement the experimentalist, C, uses the measuring apparatus, B, to measure the quantum system A and does not directly interact with A. The interaction between A and B is unitary with respect to C. For example: $U = \sum_{ij} | A(i), B(j) > < B'(i,j), A'(i)|$ where $A(i)$, $A'(i)$, $B(i)$ and $B'(i)$ are orthonormal sets of states in the respective spaces. The $B'(i,j)$ are cyclic permutations of the $B'(i)$ with respect to $j$ (chosen to make the whole evolution unitary). In the normal textbook examples $A'(i) = A(i)$. For this interaction A gains no information (makes no measurement) on the initial state of B but B gains information on A. Since the measuring apparatus must be in a known initial state for the measurement to give unambiguous information on A, the unitary transformation can be written as $U = \sum_i | A(i), B(init) > < B'(i), A'(i)|$. The state, $B'(i)$, is frequently referred to as a pointer state and depends only on the initial state of B and the interaction between A and B. The entangled wave function is: $\Psi^*(A, B) = \sum_i < A(init)| A(i) > < B'(i), A'(i)|$. The form of the unitary transformation chosen leads to an ideal measurement were a complete set of eigen functions of A are mapped uniquely to final states of B. The states $A(i)$ and $B'(i)$ must be orthogonal in their respective spaces for the mapping of the initial states of A to the final states of B to be unique.

C then performs a measurement on B, $U = \sum_i | B'(i), C(init) > < C'(i), B''(i)|$, and by self observation discovers itself to be in the definite state $< C'(k)|$. The systems B and A are then inferred to be their respective states $< B''(k)|$, and $< A'(k)|$. Relative to C, no measurement is made until C interacts with B although relative to B a measurement is made as soon as the wave functions of A and B become entangled. Since C does not interact with A, the trace over A can be done to yield the diagonal reduced density matrix of B as $\rho = \sum_i |B'(i) > | < A(init) | A(i) > |^2 < B'(i)|$. Hence already at this stage we have decoherence. The form of the unitary transformation given above assumes an ideal measurement of B by C. More generally $B'(i)$ could be replaced by a different set of orthonormal states $U_\beta = \sum_i | \beta(i), C(init) > < C'(i), \beta'(i)|$. Then the trace over the states of A or the interaction with the environment, E, discussed below, would be needed to eliminate the quantum interference terms.

C observing itself to be in a definite state is the wave function collapse with respect to C. Note that the $C'(i)$ are the C part of the Schmidt decomposition of the entangled CB system generated by the interaction with the measurement apparatus B. The wave function collapse is responsible for both the information gain and information loss in the measurement process that Rovelli noted[1]. The net information content stays the same.

One measurement gives very little information on the state of A, only its final state and that $|<A(init)|A(i)>|^2$ is non zero. If the experiment is repeated many times, the measured distribution of the $B'(i)$ gives the experimental values of $|<A(init)|A(i)>|^2$, that is the $|<A(init)|A(i)>|^2$ are the frequentist probabilities for B to be in a given state $i$. Note that there is no need to be concerned about the abstract concept of probability. Quantum mechanics simply makes predictions about the ratio of different outcomes if an experiment is repeated many times. The term probability, in this context, is just a convenient shorthand notation.

If it is known beforehand that the system is in one of a given set of eigen states, for example of angular momentum, the $A(i)$ can be chosen to match those quantum states. Then a single measurement will show which of the predetermined states that system is in. It is to the result of this type of measurement that the term fact most naturally applies. For a macroscopic system it is frequently easy to chose the $A(i)$ to approximately match (eigen) states, for example position, of the system being measured so the measurement only causes a slight perturbation and gives a uniquely determined value or fact.

All that is required for the measurement is the unbroken chain from $<A(init)|A(i)>$ to $B'(i)$ to $C'(i)$ with the value of $i$ determined by C's self observation. Decoherence plays no role. Even the Schmidt biorthonormal decomposition is only important for C and is not needed for B. C could even interact with A and perform the measurement directly. Self observation by C in the entangled state is all that is needed to collapse the wave function and effect a measurement.

4. **Interaction with the Environment**

Consider the case where after B interacts with A it interacts with the environment, E, using the interaction: $U = \sum_i |B'(i), E(init)><E'(i), B''(i)|$. The form of $U$ used here is frequently implicitly assumed, at least approximately, in discussions of decoherence. E is, in effect, performing a measurement on B. The new entangled wave function is then: $\Psi^*(A,B) = \sum_i <A(init)|A(i)><E(i), B''(i), A'(i)|$. If C does not interact with E the reduced density matrix is $\rho_E = \sum_i |A(i), B''(i)> |<A(init)|A(i)>|^2 <B''(i), A(i)|$. The $B''(i)$ become the new pointer states. The reduced density matrix is still diagonal but is in the space of A × B not just B.

If the unitary transformation for C interacting with B is $U = \sum_i |B''(i), C(init)><C'(i), B'''(i)|$ the interaction with E does not have any practical effect. Experimentalists go to great lengths to prevent the interaction of E with B from adversely affecting the experimental information obtained by B, for example by arranging that $B''(i) = B'(i)$. Both quantum and classical systems can forward information without degrading it. However if the transformation is $U_\beta = \sum_i |\beta(i), C(init)><C'(i), \beta'(i)|$ then using $\rho_E$ the probability for C being in state $j$ is

$P(j; C) = \sum_i |<\beta(j)|B''(i)>|^2 |<A(init)|A(i)>|^2 = \sum_i P(j; C, i; B) P(i; B)$. Without the decoherence due to the interaction with E there would be cross terms and quantum interference effects. Hence the interaction with E (or the trace over the states of A) is necessary to reduce the quantum probability amplitudes to quasi-classical probabilities. Classical because they propagate like classical probabilities with no interference terms. Quasi because they are based on an underlying quantum framework. Fully classical probabilities arise when the uncertainty is due entirely to the relevant interactions not being fully known or their effects not being fully calculated. See Di Biagio and Rovelli for a similar result[5] (eqns 3 and 8). In the present notation their equation is,
$P(b; C) = \sum_i P(b; C, i; B) P(i; B)$, the same as here. However they ignore the possibility that the trace over A can give decoherence.

For a more general interaction between B and E, $P(i; B)$ is the probability B is in the state $i$ in the Schmidt decomposition of the EB entangled state. To propagate information accurately the probabilities, $P(j; C, i; B)$, must be peaked near $j = i$ or equivalently $\beta(i) \approx B(i)$. Interaction with the environment is not necessary or even particularly useful, rather it is something to be minimized.

### 5. Philosophical Considerations

Relative to the quantum system C there are three types of facts or observations: Direct observations: these are the result of measurements done by C either directly or through an intermediary like B. Relevant observations or stable facts: the result of measurements by an independent system D whose reduced density matrix is diagonal relative to C. Since there are no interference terms, C can assume D's observations are contra factual definite and the probabilities will chain classically. Irrelevant observations or facts: all observations not falling into one of the two classes above. They play no role in C's description of the universe.

The classical limit has three distinct aspects. First when typical actions characterizing a system are much larger than Planck's constant the equations of motion reduce to the classical equations and the spreading of the probability distribution becomes slow. This can be shown, for example, by using the Wigner function. Secondly, due to the slow spreading of the probability distribution, the probabilities tend to stay peaked near their classical values. This, combined with the classical equations of motion, is why macroscopic systems are observed to follow classical trajectories. Finally the ubiquitousness of decoherence, due the strong interaction of macroscopic objects with the environment, will cause the reduced density matrices to be effectively diagonal. This is why macroscopic objects can be considered to follow classical trajectories even when not being observed. Contra factual definiteness, EPR realism, and the classical trajectories of the macroscopic world are emergent properties due to the three aspects of the classical limit acting in concert. The paradoxes of quantum mechanics arise when it is assumed that these emergent properties should also be true for microscopic systems.

Interpretations of quantum mechanics generally leave the predictions intact but change the unobservable components to satisfy some preconceived ideas. Being thus unconstrained by observation they have no truth value. However the construction of scientific models is guided by parsimony (Occam's razor) in addition to phenomenological adequacy.

Wave function collapse is not an observable. Consequently the wave function can be considered, without phenomenological implications, to collapse at any point in the probability chain that has contra factual definiteness, that is from the first point in that chain where the reduced density matrix is diagonal to the end of the chain where the final observer, C in the above discussion, is aware of a definite result. The trace over **either A or E** makes the relevant reduced density matrix diagonal.

The Copenhagen interpretation of quantum mechanics assumes the macroscopic measuring device causes the wave function to collapse. This is phenomenologically acceptable since at that point interaction with the environment leads to decoherence and density matrices that are effectively diagonal. Similarly, since all known conscious entities are macroscopic, assuming consciousness causes the wave function to collapse is also phenomenologically acceptable. However the only location of wave function collapse with respect to C that is consistent with the discussion in Section 2 is at the end of the chain were C interacts with the system under consideration. It is here that the time evolution is non unitary and self observation selects a unique state from those available in the Schmidt decomposition.

More generally the wave function of a given quantum system only collapses with respect to a given observer when that observer becomes entangled with that quantum system. If the wave function or state vector is simply the information a given observer has on a quantum system it is to be expected that it collapses when and only when the given observer interacts with the quantum system. The epistomological status of the wave function also eliminates the spooky action at a distance. In effect EPR realism has been given up in return for locality.

An observer (quantum system) can take as ontologically real its own state and the state, immediately after collapse, of any system it is entangled with. This is similar to and was motivated by the claim[5] of Di Biagio and Rovelli that relative facts are real but is more precise. Connected to those two types of real states are states in the quasi-classical probability chain leading to the given observer. Since these states are connected in a contra-factual definite manner to states taken to be real they can also be considered to real without fear of contradiction or internal inconsistency. The over lapping of quasi-classical probability chains leading to different observers generates the shared reality of everyday experience.

RQM with self observation collapsing the wave function provides the most parsimonious interpretation of quantum mechanics short of "shut up and calculate". There is no need for large objects, consciousness, many worlds, non locality, agents, pilot waves, mysticism, quantum Darwinism, or any other examples of the power of human imagination. The use of the Schmidt decomposition and self observation is consistent with both the letter and the spirit of RQM. In addition self observation provides an unambiguous answer to the question of when the wave function collapses. What is given up in this interpretation is contra factual definiteness (sometimes called realism) for microscopic systems. It still exists for macroscopic objects but only as an emergent property.

The interpretation of quantum mechanics presented here is fully in line with the pragmatic philosophy of C.I. Lewis. See also "Two Dogmas of Empiricism"[6] by Lewis's student W. V. Quine. Also everything here follows from using textbook quantum formalism, relational quantum mechanics[1], and Breuer's results[2]. The Schmidt biorthonormal decomposition plays a crucial role. The intrepration presented here is called modified relational quantum mechanics because the role of self observation has been added to the original relational quantum mechanics.